\documentclass[showpacs,amsmath,amssymb]{revtex4}
\usepackage{amsfonts}

\def\HH{\mathbb{H}}
\def\id{\mathrm{id}}
\def\op{\mathrm{op}}

\begin{document}
\title{On Hopf algebroid structure 
of $\kappa$-deformed Heisenberg algebra}

\author{Jerzy Lukierski}
\affiliation{Institute\;for\;Theoretical\;Physics,
University\; of\; Wroclaw
}
\author{Zoran \v{S}koda}
\affiliation{Faculty\; of\; Science,\; University\; of\; Hradec\; Kr\'alov\'{e}
}
\author{Mariusz Woronowicz}
 \affiliation{Institute\;for\;Theoretical\;Physics, University\; of\; Wroclaw
}


\begin{abstract}
The $(4+4)$-dimensional $\kappa$-deformed quantum phase space 
as well as its $(10+10)$-dimensional
covariant extension by the Lorentz sector
can be described as Heisenberg doubles: 
the $(10+10)$-dimensional quantum phase space 
is the double of $D=4$ $\kappa$-deformed Poincar\'e Hopf algebra $\HH$
and the standard $(4+4)$-dimensional space is its subalgebra generated
by $\kappa$-Minkowski coordinates $\widehat{x}_\mu$ and 
corresponding commuting momenta $\widehat{p}_\mu$.
Every Heisenberg double appears as the total
algebra of a Hopf algebroid over a base algebra which is 
in our case the coordinate sector. We exhibit the details of this 
structure, namely the corresponding right bialgebroid and the antipode
map. We rely on algebraic methods of
calculation in Majid-Ruegg bicrossproduct basis.
The target map is derived from a formula by J-H.~Lu. The coproduct
takes values in the bimodule tensor product over a base, 
what is expressed as the presence of coproduct gauge freedom.  
\end{abstract}



\maketitle

\section{Introduction}

It is often convenient to consider covariant phase spaces, that is
to include symmetries along with the space (or space-time) generators. 
Noncommutative $\kappa$-deformed Minkowski space has been complemented with
appropriate momenta into a noncommutative phase space in a number of works. 
It has been recently argued that this noncommutative phase space has
a structure of a (topological) Hopf algebroid~\cite{1,2} and the
same for general Lie algebra type phase spaces~\cite{halg}; 
the covariant structure 
is however not included in those works. 
It is known that both the standard and general $\kappa$-phase spaces
have a structure of a Heisenberg double~\cite{luknow}, 
a Hopf algebraic generalization
of a Heisenberg algebra prominently 
used in quantum group theory~\cite{Majid,heisd}.  
Furthermore, J-H. Lu~\cite{Lu} proved that 
a finite dimensional Heisenberg double has a structure of a Hopf
algebroid. Her description can be generalized to some infinite dimensional
situations, including to the $\kappa$-phase spaces; the advantage of
covariant phase space description as well as of Heisenberg double packaging
are among the main motivations for this work. 

While Hopf algebras are quantum analogues of groups~\cite{drinfeld,wor,frt}, 
Hopf algebroids are quantum analogues of groupoids. 
Indeed, algebra of functions on a group is a commutative Hopf algebra, 
while functions on a groupoid
form a commutative Hopf algebroid. To drive our expectations in physics
let us recall some example of a groupoid. Groupoid is a many object
version of a group: it consists of objects (units) and arrows between them.
Each arrow has a source and target object, we can compose two arrows,
$g$ and $f$ into $g\circ f$ if the target of $f$ is the same as the source
of $g$; each arrow has its composition inverse and the composition is
associative. The main example is a transformation (action) groupoid
which for a group $G$ acting on a manifold $M$ encodes both
space $M$ and the action in a single groupoid.
The objects are points of $M$ and 
arrows are pairs $(g,m)\in G\times M$. 
The source of $(m,g)$ is $m$ and the target of $(m,g)$ is $g\cdot m$. 
This way we know what happens to point $m$ when we act by $g$: 
the information is in the target. Thus both the space and its
symmetries are encoded in the action groupoid. Notice again that the arrows
form the product $G\times M$ of a symmetry object and the space. 

If we replace a group $G$ by Hopf algebra $\HH$ of functions on a group
and manifold $\mathcal{M}$ 
by an algebra $B$ of coordinate
functions on $\mathcal{M}$, 
we expect that $G\times M$ will be replaced by 
some sort of a tensor product $\HH\otimes B$ and the
action of $G$ on $M$ dualizes to coaction of $\HH$ on $B$. 
In a typical construction of that type, like
in this article, the tensor product is in fact a semidirect product
(in Hopf algebra literature called the smash product algebra). 
Such examples are noncommutative analogues of transformation
groupoids. In general a {\bf Hopf algebroid} over base $B$
is a $B$-{\bf bialgebroid}
with an {\it antipode} map (see Section~\ref{sec:antipode})
where $B$-bialgebroid entails the following data.
\begin{itemize}

\item The role of the quantum arrow space of a 
bialgebroid or Hopf algebroid $\mathcal{H}$ 
is taken by the {\bf total algebra}~$H$ 
and the base $B$ is the {\bf base algebra}.

\item One supplies the (dual versions of) source map $\alpha$ 
and target map $\beta$ which are now a homomorphism 
$\alpha: B\to H$ and an antihomomorphism $\beta:B\to H$ of algebras
such that their images commute in $H$:
$$
\lbrack \alpha(b),\beta(c)\rbrack = 0,\,\,\,\,b,c\in B.
$$
This way the formula $b.h.c := h\beta(b)\alpha(c)$ equips
$H$ with $B$-bimodule structure 
(this choice leads to {\it right} $B$-bialgebroid
as opposed to left bialgebroids where $b.h.c = \alpha(b)\beta(c)h$).
\item  There is a coproduct $\Delta : H\to H\otimes H$ which is however
coassociative only when projected to the equivalence classes in 
the tensor product of $B$-bimodules 
$H\otimes_B H = (H\otimes H)/\mathcal{I}_B$~\cite{brzmil,Lu,bohmHbk,Xu}
where $\mathcal{I}_B$ 
is the {\it left} ideal in $H\otimes H$  generated by all elements of the form
$$
\alpha(b)\otimes 1 - 1\otimes \alpha(b), \,\,\,\,\,\,b\in B.
$$ 
\end{itemize}
Regarding that $\mathcal{I}_B$ 
is not a 2-sided ideal~\cite{brzmil,Lu,bohmHbk,Xu} 
(it is 2-sided in some special cases, 
for instance for the undeformed Heisenberg algebra), the quotient
is not an algebra. Thus the Hopf algebraic requirement that
the effective coproduct $\Delta:H\to H\otimes_B H$ is multiplicative (that is
$\Delta(b c)=\Delta(b)\Delta(c)$) does not make sense in general.
A subtle solution to make sense of the multiplicativity of $\Delta$ 
is the following: one has to require that the image of $\Delta$ is 
within some subbimodule of $H\otimes_B H$ 
where the factorwise product is still well defined; such a subspace exists,
it is the Takeuchi product $H\times_B H\subset H\otimes_B H$,
see e.g.~\cite{takeuchi,brzmil,bohm,bohmHbk,halg}. 

Hopf algebras come not only from groups (as function algebras)
but also from Lie algebras (as universal enveloping algebras). This
means that they can also encode infinitesimal symmetries.
Similarly, Hopf algebroids can also encode the infinitesimal actions. 
This explains that the Heisenberg algebra
which entails coordinates but also their infinitesimal translation generators
is a candidate for a quantum action groupoid. We formalize this in the
language of Hopf algebroids and in the case of $\kappa$-deformations.

In Hopf-algebraic $\kappa$-deformation scheme the general
covariant $\kappa$-deformed phase space is provided by the Heisenberg
double  $\mathcal{H}=\HH\rtimes\widetilde{\HH}$ (see e.g.~\cite{Majid,heisd}),
where $\HH=U_{\kappa}(\mathfrak{g})$ describe $\kappa$-deformed Poincar\'e--Hopf
algebra \cite{13a,majidruegg,kosmas} and $\widetilde{\HH}$ is the Hopf algebra
describing dual $\kappa$-deformed quantum Poincar\'e group~\cite{zakrz}. 
Heisenberg double is a special case of 
smash (or crossed) product algebra $\HH\rtimes V$, 
where $V$ is an $\HH$-module algebra~\cite{Majid}.
In this paper we employ the general property (see e.g.~\cite{Lu}, Sect.~6) 
that Heisenberg double algebra is equipped with the Hopf algebroid structure. 
In recent literature (see e.g.~\cite{2}) the bialgebroid structures of
deformed standard quantum phase spaces $(\widehat{x}_{\mu},\widehat
{p}_{\mu})$ with $\kappa$-Minkowski space--time sector
\begin{equation}\label{eq4}
\lbrack\widehat{x}_{0},\widehat{x}_{i}]=-\frac{i}{\kappa}\widehat
{x}_{i},\qquad\lbrack\widehat{x}_{i},\widehat{x}_{j}]=0, 
\end{equation}
and commuting fourmomenta $\widehat{p}_{\mu}$ were studied by embedding
into canonical quantum phase space algebra (we put $\hslash=1$)
\begin{equation}\label{eq5}
\lbrack x_{\mu},x_{\nu}]=[p_{\mu},p_{\nu}]=0,\qquad\lbrack x_{\mu
},p_{\nu}]=i\eta_{\mu\nu}. 
\end{equation}
Relation~(\ref{eq4}) permits the following general class of realizations of
quantum phase spaces
\begin{equation}\label{eq6}
\widehat{x}_{\mu}= f_{\,\mu}^{\nu}(p) x_{\nu},\qquad\widehat{p}_{\mu}=p_{\mu}, 
\end{equation}
where $f_{\,\mu}^{\nu}(p)$ are chosen 
in consistency with relations~(\ref{eq4}), (\ref{eq5}) (Jacobi identities) 
and provide large variety of quantum
phase spaces with space--time algebra described by relations~(\ref{eq4}). 
This approach lacks structural indication how to
obtain the covariant action of $\kappa$-deformed Poincar\'e--Hopf algebra,
which is a part of full definition of quantum $\kappa$-deformed Minkowski
space.

Here the $\kappa$-deformed quantum
phase space is constructed as the Heisenberg double of $D=4$ $\kappa$-deformed
Poincar\'e--Hopf algebra; this method is first presented in~\cite{luknow}. 
Such construction contains built-in 
$\kappa$-covariance of $\kappa$-deformed quantum phase space first
observed for $\kappa$-Minkowski space--time sector in~\cite{majidruegg}. In
Majid--Ruegg basis \cite{majidruegg} 
we obtain that both $\kappa$-Poincar\'e--Hopf
algebra $\HH$ and $\kappa$-Poincar\'e group $\widetilde{\HH}$ are
described by
two dual bicrossproduct structures \cite{luknow,Maj1988,lukx,13c},
namely
\begin{equation}
\HH=U(so(1,3))\triangleright_{\!\kappa}\!\!\!\!\blacktriangleleft 
\mathcal{T}^{4}\qquad
\mathop{\longleftrightarrow}\limits^{\mathrm{duality}} \qquad
\widetilde{\HH}=\widetilde{\mathcal{T}}_{\kappa}^{4}
\triangleright_{\!\kappa}\!\!\!\!\blacktriangleleft \mathcal{L}^{6}, \label{6a}
\end{equation}
where $\mathcal{L}^{6}$ describe the functions of Abelian Lorentz parameters
$\lambda_{\mu\nu}^{\quad}$ which are dual to $U(so(3,1))$ and
$\mathcal{T}^{4}$ is the fourmomenta sector dual to the algebra
$\widetilde{\mathcal{T}}_{\kappa}^{4}$ describing noncommutative
functions of $\kappa$-deformed
Minkowski coordinates (see~(\ref{eq4})). 
The $\kappa$-Poincar\'e covariance of fourmomentum sector $\mathcal{T}^{4}$
can be derived from the bicrossproduct structure of $\HH$ (see~(\ref{6a})).
The $\kappa$-deformed Poincar\'e algebra $\HH$ acts on standard
$\kappa$-defor\-med quantum phase space 
$\mathcal{T}^{4}\rtimes\widetilde{\mathcal{T}}_{\kappa}^{4}$ in a covariant way. 
Further, the covariant action of $\HH$ on
$\mathcal{L}^{6}$ follows from the duality of $\mathcal{L}^{6}$ and
$U(so(3,1))$ algebras as well as the semidirect product of the coalgebra
sectors in $\HH$ and $\widetilde{\HH}$.

In Section 6 of~\cite{Lu}, J-H. Lu has given explicit formulas for 
a Hopf algebroid structure on a Heisenberg double of any finite dimensional
Hopf algebra, which is here replaced by 
$\kappa$-Poincar\'e Hopf algebra $\HH$ (which is $\infty$-dimensional as
a Hopf algebra, but the recipes still work). To have the formulas
explicit, in this article we calculate the details of the Hopf algebroid 
structure for the Heisenberg double 
$\mathcal{H}^{(4,4)}\equiv\mathcal{H}_{(p,x)}=
\mathcal{T}^4\rtimes\tilde{\mathcal{T}^4}$, 
where the dual Hopf algebras $\mathcal{T}^4,\tilde{\mathcal{T}}^4$
describe respectively the momentum and coordinate sectors

\begin{equation}\label{eq7}
\mathcal{T}^4\left\lbrace 
\begin{array}{l}
\lbrack\widehat{p}_{\mu},\widehat{p}_{\nu}]=0,
\\
\Delta(\widehat{p}_{i})=\widehat{p}_{i}\otimes e^{-{\frac{\widehat
{p}_{0}}{\kappa}}}+1\otimes\widehat{p}_{i},\\
\Delta(\widehat
{p}_{0})=\widehat{p}_{0}\otimes1+1\otimes\widehat{p}_{0}, 
\end{array}\right.
\end{equation}
\begin{equation}\label{7a}
\tilde{\mathcal{T}}^4\left\lbrace \begin{array}{l}
\lbrack\widehat{x}_{0},\widehat{x}_{i}]=-\frac
{i}{\kappa}\widehat{x}_{i},\qquad\lbrack\widehat{x}_{i},\widehat
{x}_{j}]=0,\\
\Delta(\widehat{x}_{\mu})=\widehat{x}_{\mu}\otimes1+1\otimes
\widehat{x}_{\mu}. 
\end{array}\right.
\end{equation}

Most earlier important examples of Hopf algebroids
over noncommutative base are also related to physics. It has been argued
out in~\cite{MackSchomerus} that low dimensional QFT-s allow for
weak Hopf algebra symmetries where weak roughly means that the coproduct 
and unit are not compatible, $\Delta(1)\neq 1\otimes 1$; usually one dispenses
with weak units passing to appropriate quotient Hopf algebras, but at the 
cost of having nonphysical zero-norm states. The data of a weak Hopf algebra 
has an equivalent description as a (special kind of a) Hopf algebroid. 
Similarly, the dynamical quantum Yang-Baxter equation (dQYB)
which describes dynamical quantum group related with Lie algebra 
$\mathfrak{g}$ 
are described~\cite{GN,EtingofNikshych}  
by a Hopf algebroid with a commutative base algebra $B$ 
dual to the Cartan subalgebra of $\mathfrak{g}$ \cite{EV}. 
It has been shown in~\cite{Xu} how to reduce dQYB 
to the ordinary Yang-Baxter equation in a Hopf algebroid framework. 
As a vector space, this algebroid is built from 
the algebraic sector of the quantum group by tensoring 
with the space encoding the dynamical parameter. 
This conceptually attractive approach 
is useful for further generalizations 
and application of twists~\cite{doninmudrov}.

All algebras in the paper are over the field $\mathbb{C}$ of
complex numbers and the opposite algebra to $B$ is denoted by $B^\op$.
We freely use the Sweedler notation for the comultiplication 
$\Delta(b) = \sum b_{(1)}\otimes b_{(2)}$ with or without summation sign.
The antipode for a Hopf algebra is denoted by $S$,
but the antipode for a Hopf algebroid by $\tau$.
We mention that in this paper we follow research partly explained in
our last publication~\cite{lukx}; however the derivation of the target map in
Section~\ref{sec:beta} is new. 

\section{$\kappa$-Poincar\'e--Hopf algebra and its dual group}

We present here the $\kappa$-Poincar\'e--Hopf algebra $\HH$ as the basic
object and the $\kappa$-Poincar\'e group as its dual Hopf algebra. Their
duality will play the major role in the next section on the Heisenberg double.

\subsection{$\protect\kappa$-Poincar\'{e}--Hopf algebra $\HH$}

In the following we use the conventions for the indices $\mu,\nu,\lambda,\sigma
=0,1,2,3$; $i,j=1,2,3$ and metric  $g_{\mu\nu}=(-1,1,1,1)$. 
We denote the $\kappa$-Poincar\'e algebra generators by $(\widehat
{p}_{\mu},\widehat{m}_{\mu\nu})$ and set $\hslash=1$.
Then the $\kappa$-Poincar\'{e}--Hopf algebra $\HH$ in bicrossproduct basis
\cite{majidruegg} has the following form 

\medskip\noindent-- \textit{algebra sector}:
\begin{equation}\label{kapp}\begin{array}{l}
\lbrack\widehat{m}_{\mu\nu},\widehat{m}_{\lambda\sigma}] =i\left(
g_{\mu\sigma}\widehat{m}_{\nu\lambda}+g_{\nu\lambda}\widehat
{m}_{\mu\sigma}-g_{\mu\lambda}\widehat{m}_{\nu\sigma}-g_{\nu\sigma
}\widehat{m}_{\mu\lambda}\right) \\
\lbrack\widehat{m}_{ij},\widehat{p}_{\mu}] =-i\left( g_{i\mu
}\widehat{p}_{j}-g_{j\mu}\widehat{p}_{i}\right)  \\
\lbrack\widehat{m}_{i0},\widehat{p}_{0}] =i\widehat{p}_{i}
,\qquad\lbrack\widehat{p}_{\mu},\widehat{p}_{\nu}]=0
\\
\lbrack\widehat{m}_{i0},\widehat{p}_{j}] =i\delta_{ij}\left( \kappa
\sinh({\frac{\widehat{p}_{0}}{\kappa}})e^{-{\frac{\widehat
{p}_{0}}{\kappa}}}+{\frac{1}{2\kappa}}\overrightarrow{\widehat
{p}}^{2}\right) -{\frac{i}{\kappa}}\widehat{p}_{i}\widehat{p}_{j}
\end{array}
\end{equation}
\noindent-- \textit{coalgebra sector}:
\begin{equation}\label{kapp2}\begin{array}{l}
\Delta(\widehat{m}_{ij})=\widehat{m}_{ij}\otimes I+I\otimes
\widehat{m}_{ij}
\\
\Delta(\widehat{m}_{k0})=\widehat{m}_{k0}\otimes e^{-{\frac
{\widehat{p}_{0}}{\kappa}}}+I\otimes\widehat{m}_{k0}+{\frac
{1}{\kappa}}\widehat{m}_{kl}\otimes\widehat{p}_{l}
 \\
\Delta(\widehat{p}_{0})=\widehat{p}_{0}\otimes I+I\otimes
\widehat{p}_{0}
\\
\Delta(\widehat{p}_{k})=\widehat{p}_{k}\otimes e^{-{\frac
{\widehat{p}_{0}}{\kappa}}}+I\otimes\widehat{p}_{k}
\end{array}
\end{equation}
\noindent-- \textit{counit:} 
\begin{equation}
\epsilon(\widehat{p}_{\mu}) = 0,
\,\,\,\,\,\,\,\,\epsilon(\widehat{m}_{\mu\nu})=0.
\end{equation}
\noindent-- \textit{Hopf algebra antipode}:
\begin{equation}\begin{array}{cc}
S(\widehat{m}_{ij})=-\widehat{m}_{ij}, & 
S(\widehat{m}_{i0})=-\widehat{m}_{i0}+\frac{3i}{2\kappa
}\widehat{p}_{i} \\
S(\widehat{p}_{i})=-e^{{\frac{\widehat{p}_{0}}{\kappa}}}\widehat
{p}_{i},& S(\widehat{p}_{0}) = -\widehat{p}_{0}.
\label{kapp3}
\end{array}
\end{equation}

\subsection{The concept of a Hopf pairing}

We say that Hopf algebra $\HH = (A,m,\Delta, S,\epsilon)$ is in duality 
with Hopf algebra  $\tilde\HH = (A^*,m^*,\Delta^*, S^*,\epsilon^*)$ 
if there is a vector space pairing
 $\langle,\rangle : A^*\otimes A\to\mathbb{C}$~\cite{Majid}
such that 
\begin{equation}\label{eq:Ha}
\begin{array}{l}
\langle c^*, a b\rangle = \langle \Delta(c^*), a\otimes b\rangle,\\
\langle  c^*\otimes d^*, \Delta(a) \rangle = \langle c^* d^*, a\rangle,
\end{array}\end{equation} 
$\langle a^*, 1_A\rangle = \epsilon^*(a^*)$,
$\langle 1_{A^*},a\rangle = \epsilon(a)$ 
and $\langle S (b^*), a\rangle = \langle b^*, S(a)\rangle$.
Any two Hopf algebras in duality act one on another. Namely,
$$
a^*\triangleright a := a_{(1)}\langle a^*, a_{(2)}\rangle,
\,\,\,\,\,\, a^*\triangleleft a := \langle a^*_{(1)},a\rangle a_{(2)}^*,
$$
Using~(\ref{eq:Ha}) one can directly verify that these actions are Hopf 
(in other words~\cite{Majid}, $A$ is a left $\tilde\HH$-module algebra 
and $A^*$ is a right $\HH$-module algebra), that is
$$\begin{array}{lcl}
a^*\triangleright (a b)& = &(a_{(1)}^*\triangleright a)
(a_{(2)}^*\triangleright b)
\\
&=& a_{(1)}\langle a^*_{(1)},a_{(2)}\rangle b_{(1)} \langle a_{(2)}^*,b_{(2)}\rangle
\end{array}$$

\subsection{$\protect\kappa$-Poincar\'{e} quantum group $\widetilde
{\HH}$}

We introduce the $\protect\kappa$-Poincar\'{e} quantum group as the dual
vector space $\tilde\HH$ to $\HH$ with generators 
$\widehat{p}_{\nu},\widehat{m}_{\lambda\mu}$ 
via the following canonical duality relations
\begin{equation}\label{dual}\begin{array}{c}
<\widehat{x}^{\mu},\widehat{p}_{\nu}>=i\delta_{\nu}^{\mu}\\
<\widehat{\lambda}{^{\mu}}_{\nu},\widehat{m}_{\lambda\rho}>=
i(\delta_{\lambda}^{\mu}g_{\nu\rho}-\delta_{\rho}^{\mu}g_{\nu\lambda}
- \delta_\lambda^\nu g_{\mu\rho}+ \delta_\rho^\nu g_{\mu\lambda})
\\
\langle \widehat{x}^\mu,\widehat{m}_{\lambda\rho}\rangle = 
\langle \widehat\lambda{^\mu}_\nu, \widehat{p}_\sigma\rangle  = 0,\quad
\langle \widehat\lambda{^\mu}_\nu, I\rangle = \delta^\mu_\nu
\end{array}
\end{equation}
The pairing for vector spaces has to be given for vector space basis
but in the case of Hopf algebra duality it is enough to specify the
pairing just on the algebra generators. Indeed, for products of generators 
we can use~(\ref{eq:Ha}).

From~(\ref{dual}) 
we obtain the commutation relations defining $\kappa$-Poincar\'{e} group
\cite{zakrz,13c} in the following form

\medskip\noindent-- \textit{algebra sector}:
\begin{equation}\begin{array}{l}
\lbrack\widehat{x}^{\mu},\widehat{x}^{\nu}]={\frac{i}{\kappa
}}(\delta_{0}^{\mu}\widehat{x}^{\nu}-\delta_{0}^{\nu}\widehat
{x}^{\mu
})
,\,\,\,\,\,\,
\lbrack\widehat{\lambda}_{\nu}^{\mu},\widehat{\lambda
}_{\beta
}^{\alpha}]=0 \\
\lbrack\widehat{\lambda}_{\nu}^{\mu},\widehat{x}^{\lambda}]=
-{\frac{i}{\kappa}}\left( (\widehat{\lambda}_{0}^{\mu}-\delta
_{0}^{\mu})\widehat{\lambda}_{\nu}^{\lambda}+(\widehat{\lambda
}_{\nu}^{0}-\delta
_{\nu}^{0})g^{\mu\lambda}\right)
\label{grup}
\end{array}\end{equation}
-- \textit{coalgebra sector}:
\begin{equation}\label{eq:xlcop}\begin{array}{l}
\Delta(\widehat{x}^{\mu})=\widehat{\lambda}{^{\mu}}_{\rho
}\otimes
\widehat{x}^{\rho}+\widehat{x}^{\mu}\otimes I
 \\
\Delta(\widehat{\lambda}{^{\mu}}_{\nu})=\widehat{\lambda
}{^{\mu}}_{\rho}\otimes\widehat{\lambda}{^{\rho}}_{\nu}
\end{array}\end{equation}
-- \textit{ counit $\epsilon$ and antipode $S$}:
\begin{equation}\label{ant}\begin{array}{lll}
\epsilon(\widehat
{\lambda}{^{\mu}}_{\nu})=\delta{^{\mu}}_{\nu} &\,\,\,\,\,\,\,\,& 
S(\widehat{\lambda}{^{\mu}}_{\nu})
= g^{\mu\rho}\widehat{\lambda}_{\ \rho}^{\sigma}g_{\nu\sigma} 
= \widehat{\lambda}_{\nu}{^{\mu}} \\
\epsilon(\widehat{x}^{\mu}) =0 && 
S(\widehat{x}^{\mu})=-\widehat{\lambda}_{\nu}{^{\mu}}
\widehat{x}^{\nu}
\end{array}\end{equation}

In the Heisenberg double algebra
$\mathcal{H}^{(10,10)}=\HH\rtimes\widetilde{\HH}$ the commutation
relations~(\ref{kapp}) and~(\ref{grup}) are supplemented by the following
relations obtained from~(\ref{dual}),
(\ref{kapp2}) and~(\ref{eq:xlcop})

The generalized covariant $\kappa$-deformed phase space is described by
sets of commutators~(\ref{kapp}),
(\ref{grup}), (\ref{cross}) and~(\ref{cross1}). The
coproducts~(\ref{kapp2}) and~(\ref{eq:xlcop}) realize the coalgebraic
homomorphism of relations~(\ref{kapp}) and~(\ref{grup}), but the
relations~(\ref{cross}),(\ref{cross1}) 
will be mapped into the coalgebra only in the bialgebroid
framework. 

The Hopf subalgebra $\tilde{\mathcal{T}}^4_\kappa = U_\kappa(g)$ 
($\kappa$-Minkowski space-time sector) 
is the subalgebra generated by the 4 generators $\widehat{x}_\mu$ only. 

\section{Heisenberg double}

Heisenberg double is a construction of an associative algebra $A\rtimes A^*$
(the notation suggests that it is a special case of a smash product
algebra)
containing a Hopf algebra $\HH$ and its dual Hopf algebra $\tilde\HH$
as the analogues of the coordinate 
and momentum sectors within the standard Heisenberg algebra. 
It is the tensor product vector space $A\otimes A^*$ with the nontrivial 
algebra structure given by the cross relations.

\subsection{Cross relations}

The two tensor factors do not commute, but satisfy the cross relations
\begin{equation}\label{eq11}
(1\otimes a^\ast)(a \otimes 1) = 
(a^\ast_{(1)}\triangleright a)\otimes a_{(2)}^\ast
 = a_{(1)} <a_{(1)}^{\ast},a_{(2)}>\otimes\, a_{(2)}^\ast,
\end{equation}
while $(a\otimes 1)(1\otimes a^*) = a\otimes a^*$. Regarding that 
we know that $A$ is the first and $A^*$ the second tensor factor
in practice we may concatenate and skip the $\otimes$ sign.  
Thus $a\otimes a^* = a a^*$ and~(\ref{eq11}) reads $a^\ast a =  
a_{(1)} <a_{(2)},a_{(1)}^{\ast}> a_{(2)}^\ast$.

Using coproducts~(\ref{kapp2}), (\ref{eq:xlcop}) 
we calculate~\cite{meljcorr}
the cross relations~(\ref{eq11})
\begin{equation}\label{cross}\begin{array}{cc}
\lbrack\widehat{p}_{k},\widehat{x}_{l}]=-i\delta_{kl},
&
\lbrack\widehat{p}_{0},\widehat{x}_{0}]= i  \\
\lbrack\widehat{p}_{k},\widehat{x}_{0}]=-{\frac{i}{\kappa
}}\widehat{p}_{k}, & \lbrack\widehat{p}_{0},\widehat
{x}_{l}]=0,
\\
 \lbrack\widehat{p}_{\mu},\widehat{\lambda}_{\rho}^{\lambda
}]=0, &
\end{array}\end{equation}
\begin{equation}\label{cross1}\begin{array}{l}
\lbrack\widehat{m}_{\lambda\rho},\widehat{\lambda}_{\nu}^{\mu}]=
i\left(\delta_{\rho}^{\mu}\widehat{\lambda}_{\lambda\nu}
-\delta_{\lambda}^{\mu}\widehat{\lambda}_{\rho\nu}
+\delta_{\lambda}^{\nu}\widehat{\lambda}_{\rho\mu}
-\delta_{\rho}^{\nu}\widehat{\lambda}_{\lambda\mu} \right),\\
\lbrack\widehat{m}_{\lambda\rho},\widehat{x}^{\mu}]=
i\left(\delta_{\rho}^{\mu}\widehat{x}_{\lambda}-\delta_{\lambda}^{\mu}\widehat
{x}_{\rho}\right) +{\frac{i}{\kappa}}\left( \delta_{\rho
}^{0}\widehat{m}{_{\lambda}}^{\mu}-\delta_{\lambda}^{0}\widehat
{m}{_{\rho}}^{\mu}\right)
\end{array}\end{equation}
where $\widehat{m}{_{\lambda}}^{\mu}=g^{\mu\rho}\widehat
{m}{_{\lambda\rho}}$ and $\widehat{m}{^{\mu}}_{\lambda}=g^{\mu
\rho}\widehat{m}{_{\rho\lambda}}$.

The general covariant $\protect\kappa$-deformed quantum phase space
$\mathcal{H}^{(10,10)}$ is generated by $\HH$, $\tilde\HH$
with the above cross relations. The standard $\kappa$-deformed 
phase space $\mathcal{H}^{(4,4)}\subset\mathcal{H}^{(10,10)}$ 
is its subalgebra generated by $\widehat{x}^\mu$ and $\widehat{p}_\nu$ only.
The quotient of $\mathcal{H}^{(10,10)}$ 
by the relations $\widehat\lambda^\mu_{\ \nu} = \delta^\mu_\nu$ 
is the covariant $\kappa$-deformed DSR algebra.
The subalgebra with generators $\widehat{x}^\mu$, $\widehat{p}_\nu$ and
$\widehat{m}_{\lambda\sigma}$ is dual to the $\kappa$-deformed DSR algebra.

\subsection{Heisenberg double as a right bialgebroid}

The Heisenberg double $\mathcal{H}$ 
of any finite-dimensional Hopf algebra $\HH$ is shown in 
the Section 6 of~\cite{Lu} to have the structure of 
a Hopf $A^*$-algebroid where the base $B:=A^*$ 
is the underlying algebra of $\HH$. Our Hopf algebroid
is an instance of an $\infty$-dimensional version of that algebroid. 
We shall neglect the mathematical questions 
of completions (see e.g.~\cite{halg}).

The recipes of J-H. Lu~\cite{Lu} are the following. 
The source $\alpha(a^\ast) = 1\otimes a^\ast\in H = A\rtimes A^*$ 
for $a^*\in A^*$, (usually we identify 
$A^\ast\cong {\mathbb C}\otimes A^\ast\hookrightarrow A\rtimes A^\ast$).
For the target Lu suggests a formula involving the canonical 
element. In Section~6 of Lu~\cite{Lu} 
a formula for the target map $\beta$ of the left bialgebroid structure
on a Heisenberg double of finite dimensional Hopf algebra is written out; 
this formula involves the {\it canonical element} in $A\otimes A^*$.  Given 
a vector space basis $\{ f^I\}_I$ of $A$ 
and the dual basis $\{h_I\}_I$ of $A^*$ 
characterized by $\langle f^I,h_J\rangle = \delta^I_J$,
the canonical element is $\sum_I f^I\otimes h_I$. If $A$ is
infinite dimensional then the dimension of $A^*$ is of even bigger cardinality
than that of $A$, hence it requires some care to make sense of the canonical
element. Still, the dual vectors $f^I$ are well defined
and, in cases like ours, the infinite sum converges in 
a (formal) completion $A\widehat\otimes A^*$. 
M. Stoji\'c has pointed to us
a {\it right} bialgebroid version of Lu's formula for $\beta$, namely
\begin{equation}\label{eq:LuStojic}
\beta(t) = \sum_{I,J} f^J S^{-1} f^I \otimes h_I h h_J \in A\rtimes A^\ast
\end{equation}
where $\sum_I f^I \otimes h_I$ is the canonical element.

For the effective comultiplication $\Delta : H\to H\otimes_B H$  
(where $B$ is the base) J-H.~Lu~\cite{Lu} defines 
\begin{equation}\label{eq:comLu}
\Delta(a\otimes b^\ast) = \sum (a_{(1)} \otimes 1)\otimes_B (a_{(2)}\otimes b^\ast),
\,\,\,\,\,\,\,\,\,\,\,\,\,a\in A, b\in A^\ast,
\end{equation}
where $\Delta_{\HH}(a) = \sum a_{(1)}\otimes a_{(2)}$ is the coproduct in
the Hopf algebra $\HH$. Of course, in~(\ref{eq:LuStojic}) 
we can simply write $a_{(1)}\otimes a_{(2)} b^\ast$.

In the case of $\mathcal{H}^{(4,4)}$, this coproduct $\Delta$ 
is given on the generators by the formulae
\begin{equation}\label{copH44}\begin{array}{ll}
\Delta(\widehat{x}^\mu) = 1\otimes\widehat{x}^\mu &\\
\Delta(\widehat{p}_k) = \widehat{p}_k\otimes e^{\-\frac{\widehat{p}_0}\kappa}
+ I\otimes\widehat{p}_k,&
\Delta(\widehat{p}_0) = \widehat{p}_0\otimes I + I\otimes\widehat{p}_0.
\end{array}\end{equation}

\subsection{Counit and Fock actions}

The property of the counit $\epsilon:H\to B$ 
which one wants to preserve from the Hopf algebra case 
is its relation to the coproduct, that is
\begin{equation}\label{eq:eps}
(\epsilon\otimes_B \id)\Delta = (\id\otimes_B\epsilon)\Delta = \id
\end{equation}
This requirement makes sense in view of the canonical identification 
of the $B$-bimodules $B\otimes_B H\cong H\otimes_B B \cong H$.
Taking into account the $B$-bimodule structure on $H$ (used in the 
bimodule tensor product over $B=A^\ast$), this amounts to
$h_{(1)}\alpha(\epsilon(h_{(2)})) = h = h_{(2)}\beta(\epsilon(h_{(1)}))$.
However, it is generally not possible to force the counit to be an algebra
homomorphism. Instead it satisfies the weaker properties
\cite{brzmil,bohmHbk}
\begin{equation}
\epsilon\left(\alpha(\epsilon(h))h'\right) =
\epsilon(h h') = \epsilon\left(\beta(\epsilon(h)) h'\right)
\end{equation}
In the case of the Heisenberg double $A\rtimes A^\ast$ define
\begin{equation}
\epsilon(a\otimes b^*) = \epsilon_{A}(a) b^*,\,\,\,\,\,\,\,\,\,\,\,\,\,
a\otimes b^*\in A\rtimes A^\ast
\end{equation}
Then 
$(f,h)\mapsto f\blacktriangleleft h = \epsilon(\alpha(f)h)$ 
is a right action, where
$f\in\widetilde{\mathcal{T}_{\kappa}^{4}}$ and
$h,h'\in\mathcal{H}^{(4,4)}=\mathcal{T}^{4}\rtimes\widetilde{\mathcal
{T}_{\kappa}^{4}}$. 
This action should be viewed as a deformed Fock space where the unit
$1 =: \langle 0 |$ is the right Fock vacuum. 
Indeed, $\langle 0 | \widehat{p}_\mu = 0$ 
and $\langle 0 | \widehat{x}^\nu = \widehat{x}^\nu$ 
and the usual normal ordering
and commuting procedures for the evaluation of long expressions in coordinates
and momenta on the vacuum apply.

The counit $\epsilon$ satisfies defining equations~(\ref{eq:eps})
and on the generators of $\mathcal{H}^{(4,4)}$ is given by
\begin{equation}
\epsilon(\widehat{x}_{\mu})=\widehat{x}_{\mu},\qquad\epsilon
(\widehat{p}_{\mu})=0,\qquad\epsilon(1)=1.
\end{equation}

In our case, the identity $\epsilon(\alpha(\widehat{x}_{\mu}))=\epsilon
(\beta(\widehat{x}_{\mu}))=\widehat{x}_\mu$ holds. 

\subsection{Calculating the target map}\label{sec:beta}

To apply the formula~(\ref{eq:LuStojic}) for $\beta$ 
to the $(4+4)$-dimensional $\kappa$-phase space we first 
compute the canonical element. This is easier in 
the undeformed case; the results is then used to compute the deformed
case. Then the monomials $x^I$ in $x^\mu$ form a basis,
and the monomials $\widehat{p}_I$ in momentum operators $\widehat{p}_\mu$
are the dual elements, up to proportionality constants, 
namely by~(\ref{dual}) and~(\ref{eq:Ha}) we obtain 
$$
\langle x^J, \widehat{p}_I  \rangle_0 = i^{|I|} I! 
$$ 
where $I = (i_0,i_1,i_2,i_3)$ is a multiindex, $|I| = -i_0+i_1+i_2+i_3$ 
and $I! = i_0! i_1! i_2! i_3!$. The canonical element is therefore
\begin{equation}\label{eq:canexp}
\sum_{i_0,i_1,i_2,i_3=0}^\infty \frac{i^{|I|}}{I!}\widehat{p}_I\otimes x^I =
\exp(i\sum_{\mu = 0}^3\widehat{p}_\mu \otimes x^\mu).
\end{equation}
There is now an isomorphism $\xi$ of coalgebras~\cite{heisd} 
from the undeformed coordinate algebra (polynomials in $x^\mu$) 
to the deformed coordinate algebra (polynomials in $\widehat{x}^\mu$). 
The inverse $\xi^{-1}$ of this isomorphism can be computed 
from the commutation relations as follows. 
In the right bialgebroid case, $\widehat{x}^\mu$ is realized as the operator
$- i [\widehat{x}^\mu, \widehat{p}_\sigma] x^\sigma$ where 
$\widehat{p}_\sigma = -i\partial_\sigma$ and the operators act to the right.
Then, an arbitrary expression $h$ in $\widehat{x}^\mu$ acts from the right 
as a differential operator 
to (the right-hand version of) the Fock vacuum, the result is $\xi^{-1}(h)
= \langle 0 | h_{\mathrm{oper}}$. 
In our case, 
$$
\widehat{x}_k = x_k,\,\,\,\,\,\,\,\,\,\,  
\widehat{x}_0 = x_0 - \frac{i}{\kappa}\sum_{j = 1}^3\partial_j x_j.
$$
Using this, one can calculate that
\begin{equation}\label{eq:expord}
\langle 0 | \exp(-i p_0 \widehat{x}_0)\exp(i p_j \widehat{x}_j) 
= \exp(-i p_0 x_0 + i p_j x_j)
\end{equation}
where $p_0, p_j$ are just numbers -- not operators and the summation over $i$, 
$\sum_{j=1}^3 p_i x_i$ is understood.

In fact we should find some functions $F^\mu(k_0,k_1,k_2,k_3)$ so that
$$
\langle 0 | \exp(i F^\mu(k) \widehat{x}_\mu) 
= \exp(-i k_0 x_0 + i k_j x_j)
$$
but this involves more involved calculations of functions $F^\mu$.
We instead take a simpler looking {\it product} of exponentials, 
on the expense of need
for further care of the ordering in the remaining computation below. 

The crucial observation~\cite{heisd} 
is now that the deformed pairing $\langle, \rangle$
can be expressed in terms of the undeformed pairing $\langle, \rangle_0$ 
and the map $\xi$, 
$$
\langle \xi(x^J),\widehat{p}_I \rangle =
\langle  x^J,\widehat{p}_I\rangle_0.
$$
Regarding that $\xi$ is an isomorphism of vector spaces, this means that
it is wise to take as basis $\xi(x^J)$ and preserve the dual basis and we
get the canonical element 
$\sum_I \frac{(-i)^{|I|}}{I!}\widehat{p}_I\otimes \xi(x^I)
= (1\otimes \xi)\exp(-i \widehat{p}_\mu \otimes x^\mu) 
= \exp(+i \widehat{p}_0\otimes
\widehat{x}_0)\exp(-i\widehat{p}_i\widehat{x}_i)$.

Here we used that the first tensor product commutes with the second,
hence behaving as formal commuting variable in the calculation above.

Note that the equality of the sums $\sum_I S^{-1} f^I\otimes h_I = \sum_I f^I\otimes S^{-1} h_I$. Indeed, the antipode $S$ of is an isomorphism as well and
$\langle S f, S^{-1} h\rangle 
= \langle f, S S^{-1} h\rangle =\langle f, h\rangle$ 
because the pairing is Hopf~(\ref{eq:Ha}). 
Therefore $\tilde{f}^I = S f^I$ also form a basis
and $\tilde{h}_I = S^{-1} h_I$ are the dual vectors;
regarding that the canonical element does not depend 
on the choice of basis, 
$\sum_I S^{-1} f^I\otimes h_I = \sum_I S^{-1} S f^I\otimes S^{-1} h_I =
\sum_I f^I \otimes S^{-1} h_I$.
In our case, $h_I$ are the elements in the enveloping algebra
and the antipode contributes to the sign, 
thus for $\sum_I f^I \otimes S^{-1} h_I$
we get the same group-like exponential~(\ref{eq:canexp}) 
with the minus sign, that is its multiplicative inverse.
Noting that $A$ is in our case commutative, the order of
$f_I$ and $f_J$ can interchange in the Lu formula~(\ref{eq:LuStojic}) 
and we obtain
\begin{equation}\label{eq:calcmain}
\beta(h) = 
\zeta\left(e^{i \widehat{p}_j\otimes \widehat{x}_j} e^{-i \widehat{p}_0\otimes \widehat{x}_0}
(1\otimes h) 
e^{i\widehat{p}_0 \otimes \widehat{x}_0} e^{-i \widehat{p}_j\otimes \widehat{x}_j}\right)
\end{equation}
where $\zeta$ is the isomorphism of vector spaces
sending the tensor product algebra 
$A\otimes A^\ast = \mathcal{T}\otimes \tilde{\mathcal{T}}^4$ 
to the smash product algebra $\mathcal{T}\rtimes \tilde{\mathcal{T}}^4$. 
This isomorphism is the identity on the vector space level 
but not a homomorphism of algebras, hence the expressions in terms of 
products of algebra generators may look different.

We want to apply this formula~(\ref{eq:calcmain}) to the generators 
$\widehat{x}_0$ and $\widehat{x}_i$;
to this end first we apply the conjugation 
with the inner exponentials and then with the outer exponentials. 
Using ad-expansion $e^X Y e^{-X} = e^{\mathrm{ad}\,X}(Y)$ 
and inductive calculations with the exponential series, 
in the case of $h = \widehat{x}_i$ the inner conjugation 
in~(\ref{eq:calcmain}) gives 
$\exp(-i \widehat{p}_0\otimes\widehat{x}_0)(1\otimes\widehat{x}_i)\exp(i \widehat{p}_0\otimes\widehat{x}_0) = \exp(-\frac{\widehat{p}_0}{\kappa})\otimes \widehat{x}_i$ 
and the outer conjugation leaves the result intact,
hence after applying $\zeta$ we obtain
\begin{equation}\label{eq:betaxi}
\beta(\widehat{x}_i) = \exp(-\frac{\widehat{p}_0}{\kappa})\widehat{x}_i.
\end{equation}
For $h = \widehat{x}_0$ the inner conjugation 
is not affecting $(1\otimes h)$ and the outer conjugation gives
\begin{equation}\label{eq:betax0}
\beta(\widehat{x}_0) 
= \zeta\left(e^{i \widehat{p}_k\otimes\widehat{x}_k} 
(1\otimes\widehat{x}_0)e^{- i \widehat{p}_k\otimes \widehat{x}_k}\right) 
= \zeta\left(1\otimes\widehat{x}_0 
      - \frac{1}{\kappa}\widehat{p}_k\otimes\widehat{x}_k\right)
= \widehat{x}_0 - \frac{1}{\kappa}\widehat{p}_k\widehat{x}_k.
\end{equation}
\subsection{The coproduct gauge freedom}

If we view the coproduct as an algebra map 
$\Delta:H\to H\otimes_{\mathbb C} H$ then
it is coassociative only up to elements in certain subspace of 
$H\otimes_{\mathbb C} H\otimes_{\mathbb C}H$ and $\Delta$ itself
can also be redefined up to elements in 
$\mathcal{I}_B\subset H\otimes_{\mathbb C} H$. 
This freedom of coproducts is parametrized by the coproduct gauges. 
If we consider coproduct as means to build the realizations of algebra 
$H$ on tensor spaces,
which corresponds in the physical context 
to the Fock space realizations and multiparticle states,
various possibilities of such tensoring depending on 
the physical choice of particular
dynamical model are possible (see also Section~\ref{sec:disc}). 
In such a framework the gauge-invariant object is the algebra $H$, 
and its tensorial realizations are gauge-dependent, different for
various dynamical models with Fock-like extensions.
The Hopf algebroid structure in mathematical sense 
deals with strictly coassociative coproduct, 
which we may call {\it effective}, 
where the freedom in the left ideal $\mathcal{I}_B$ is
quotiented out~\cite{brzmil,Xu,Lu}, hence $\Delta: H\to H\otimes_B H = 
(H\otimes_{\mathbb C} H)/{\mathcal I}_B$. 

Let us compute the ideal $\mathcal{I}_B$ 
using $\beta$ from~(\ref{eq:betaxi}) and~(\ref{eq:betax0}):
\begin{equation}\label{eq:IBx}\begin{array}{l}
\alpha(\widehat{x}_i)\otimes 1 - 1\otimes\beta(\widehat{x}_i)
= \widehat{x}_i\otimes 1 - 1\otimes e^{-\frac{\widehat{p}_0}{\kappa}}\widehat{x}_i,
\\
\alpha(\widehat{x}_i)\otimes 1 - 1\otimes\beta(\widehat{x}_i)
= \widehat{x}_i\otimes 1 - 1\otimes\widehat{x}_0 + 
1\otimes\frac{1}{\kappa}\widehat{p}_k\widehat{x}_k.
\end{array}\end{equation}
The expressions on the right hand side of equations~(\ref{eq:IBx}) generate
$\mathcal{I}_B$ as a left ideal.

We can arrive to the same ideal or its parts by considering coproduct
freedom using elementary arguments with tensors.
Thus, if $\Delta(\widehat{x}^\mu) = 1\otimes \widehat{x}^\mu$
then any other homomorphism $\tilde\Delta$ which is of the form
\begin{equation}
\widetilde{\Delta}(\widehat{x}_{\mu})=\Delta(\widehat{x}_{\mu})+\Lambda_{\mu}(\widehat{x},\widehat{p})=\widehat{x}_{\rho}\otimes
\theta_{\mu}^{\rho}(\widehat{p}), \label{mcop}
\end{equation}
where $\Delta(\widehat{x}^{\mu})=1\otimes\widehat{x}^{\mu}$ and
$\theta_{\mu}^{\rho}(\widehat{p})$ is the tensor to be determined satisfies

\begin{eqnarray}
\lbrack\Delta(\widehat{x}_{[\mu}),\Lambda_{\nu]}]+
[\Lambda_{\mu},\Lambda_{\nu}] =C_{\mu\nu}^{(\kappa)\rho}\Lambda_{\rho},
\label{r1} 
\\
\lbrack\Delta(\widehat{p}_{\mu}),\Lambda_{\nu}] =0. \label{r2}
\end{eqnarray}
where $C_{\mu\nu}^{(\kappa)\rho}=\frac{1}{\kappa}(\delta
_{\mu}^{0}\eta_{\nu}^{\rho}-\delta_{\nu}^{0}\eta_{\mu}^{\ \rho})$ are the structure 
constants, $[\widehat{x}_{\mu},\widehat{x}_{\nu}]=C_{\mu\nu}^{(\kappa
)\rho}\widehat{x}_{\rho}$.

The relations~(\ref{r1})--(\ref{r2}) are required if the transformation
$\Delta(\widehat{x}_{\mu})\longrightarrow\widetilde{\Delta}(\widehat
{x}_{\mu})$ is todescribe the coproduct gauge. Postulating that
$(\widetilde{\Delta}(\widehat{x}^{\mu}),\Delta(\widehat{p}_{\nu}))$
satisfies the
quantum phase space algebra relations 
which $\widehat{x}^\mu,\widehat{p}_\nu$ do 
one algebraically derives
the conditions fixing the tensor $\theta_{\mu}^{\rho}(\widehat{p})$, namely 
\begin{equation}\label{koa}
\widetilde{\Delta}(\widehat{x}_{i})=\widehat{x}_{i}\otimes e^{\frac
{\widehat{p}_{0}}{\kappa}},\quad
\widetilde{\Delta}(\widehat{x}_{0})=\widehat{x}_{0}\otimes1+\frac
{1}{\kappa}\widehat{x}_{i}\otimes e^{\frac{\widehat{p}_{0}}{\kappa
}}\widehat{p}_{i}. 
\end{equation}
As follows from~(\ref{mcop}) one gets
\begin{equation}\label{rr}\begin{array}{l}
\Lambda_{i} =\widehat{x}_{i}\otimes e^{\frac{\widehat{p}_{0}}{\kappa
}}-1\otimes\widehat{x}_{i}, \\
\Lambda_{0} =\widehat{x}_{0}\otimes1-1\otimes\widehat{x}_{0}+\frac
{1}{\kappa}\widehat{x}_{i}\otimes e^{\frac{\widehat{p}_{0}}{\kappa
}}\widehat{p}_{i}.
\end{array}\end{equation}

Thus we obtained that $\Lambda$-s are in the ideal $\mathcal{I}_B$,
namely they are of the form
\begin{equation}\begin{array}{l}
\Lambda_\mu(\widehat{x},\widehat{p}) = (1\otimes\theta^\rho_\mu)(\alpha(\widehat{x}_\rho)\otimes 1 - 1\otimes\beta(\widehat{x}_\rho))\in\mathcal{I}_B
\end{array}\end{equation}
Similar tensors $R_{\mu}=\widehat{x}_{\mu}\otimes1-
\tilde\theta_{\mu}^{\rho}(\widehat{p})\otimes\widehat{x}_{\rho}$,
where $\tilde\theta_{\mu}^\nu$
is the matrix inverse to $\theta_{\lambda}^\rho$ (see~(\ref{mcop})),
have been introduced in~\cite{1}
for the canonical twisted Heisenberg algebra and considered in~\cite{2}
for the $\kappa$-deformed quantum phase space generated by
the $\kappa$-deformed Poincar\'e--Hopf algebra
with classical Poincar\'e algebra sector. Note that
our $\widehat{x}^\mu$ corresponds to $\hat{y}^\mu$ from~\cite{1}
rather than their $\hat{x}^\mu$ which is a more convenient generator for 
the description of the {\it left} bialgebroid structure.

Analogous analysis as for $\Lambda_\mu$ 
can be made for higher order tensors~\cite{lukx}. 

\section{Antipode}\label{sec:antipode}

Hopf algebroid is a bialgebroid with an antipode which is
a linear map $\tau:H\to H$. 
In bialgebroids, the source $\alpha$ and 
target $\beta$ together play the role of the unit and it is natural
to require $m(\tau\otimes_B\id)\Delta=\alpha\epsilon$ and 
$m(\id\otimes_B\id)\Delta=\beta\epsilon$ as an extension
of the axioms for the antipode in Hopf algebra theory. 
The first of the two equations is however problematic: 
$m(\tau\otimes_B\id)(\mathcal{I}_B)\neq 0$ in general,
hence the right hand side 
with effective $\Delta$ being defined up to gauge in $\mathcal{I}_B$ 
is not well defined; only a subclass of gauges could satisfy this equation. 
J-H. Lu~\cite{Lu} 
avoids this by restricing to a subclass of gauges; in fact she
introduces a linear map $\gamma:H\otimes_B H\to H\otimes H$
which is a section of the projection $H\otimes H\to H\otimes_B H$,
so that $\Delta_\gamma = \gamma\circ\Delta$ is a choice of gauge. Her
axioms (here in the {\it right} bialgebroid version) are thus nonsymmetric:
\begin{eqnarray}
\tau\beta =\alpha, \label{f1}\\  
m(\tau\otimes_{\mathbb C}\mathrm{id})(\gamma\circ\Delta) 
=\alpha\epsilon, \label{f22} 
\\
m(\mathrm{id}\otimes_B\tau)\Delta =\beta\epsilon\tau.
\label{f33}
\end{eqnarray}
Now the
left hand side of~(\ref{f33}) does not depend on coproduct gauge for $\Delta$.
An approach by G. B\"ohm~\cite{bohmHbk} is symmetric: 
she supplies both a right $B$-bialgebroid $\mathcal{H}_R$ 
and a left $B^\op$-bialgebroid $\mathcal{H}_L$ structure 
on the same total algebra $\mathcal{H} = \mathcal{H}_R = \mathcal{H}_L$
and the antipode $\tau:\mathcal{H}\to\mathcal{H}$ is a linear antiautomorphism
satisfying a number of axioms involving both left and right bialgebroid
structures; the coproducts of $\mathcal{H}_L$ and $\mathcal{H}_R$ are
also compatible as well as the counit, source and target data for the
two bialgebroids. 
In \cite{4} a subalgebra $\mathcal{B}$ in $H\otimes_{\mathbb C} H$ is
singled out within which all gauge choices satisfy~(\ref{f22}), i.e. $\gamma\circ\Delta$ can be replaced by any $\tilde\Delta$ 
landing within $\mathcal{B}$. The
latter approach has been axiomatized in~\cite{twosha}.

To compute the antipode, we can either again follow
the Heisenberg double formulas for $\tau$ in Section~6 of~\cite{Lu} 
which, like for $\beta$,
also involve the canonical elements. 
But this recipe boils down to the following direct reasoning. 
For the momentum sector $\mathcal{T}^{4}$,
$\tau$ agrees with the Hopf-algebraic antipode:
$\tau(\widehat{p}_0) = S(\widehat{p}_{0})=-\widehat{p}_{0}$ and
$\tau(\widehat{p}_i) =
S(\widehat{p}_{i}) = -e^{{\frac{\widehat{p}_{0}}{\kappa}}}\widehat{p}_{i}$
(see~(\ref{kapp3})).
For the coordinate sector, formulas for $\tau$ are forced by the equations
$\tau(\beta(\widehat{x}_\mu)) = \alpha(\widehat{x}_\mu) = \widehat
{x}_\mu$
which, using that $\tau$ is an antihomomorphism of algebras
and inserting its values on $\widehat{p}_i$, yield
\begin{eqnarray}
\tau(\widehat{x}_{i}) =e^{-\frac{\widehat{p}_{0}}{\kappa}}\widehat
{x}_{i}=\tau^{-1}(\widehat{x}_i)=\beta(\widehat{x}_{i}), \\
\tau(\widehat{x}_{0}) =\widehat{x}_{0}-\frac{1}{\kappa}\widehat
{x}_{i}\widehat{p}_{i}=\beta(\widehat{x}_{0})-\frac{3i}{\kappa},
\end{eqnarray}
\begin{equation}
\tau^{2}(\widehat{p}_{\mu})=\widehat{p}_{\mu},\,\,\,\,\,\,\,\,\,\,\,\,
\,
\tau^{2}(\widehat{x}_i)=\widehat{x}_i,\,\,\,\,\,\,\,\,\,\,\,\,\,
\tau^{2}(\widehat{x}_0)=\widehat{x}_0 - \frac{3i}{\kappa}.
\end{equation}

\section{Discussion and Outlook}\label{sec:disc}
The general aim of this paper is to promote the application of Hopf algebroid  structure to the description of quantum--deformed phase spaces. We consider here the algebraic model of phase spaces constructed as Heisenberg doubles or equivalently smash products of  Drinfeld quantum enveloping algebra $\mathbb{H}$ acting upon dual quantum group $\widetilde{\mathbb{H}}$.
In bialgebroid framework an important difference with bialgebra is the generalization of unity to the noncommutative base algebra. 
If we consider the quantum-deformed phase spaces as Heisenberg doubles the base algebra is provided by the whole algebraic sector of quantum group $\widetilde{\mathbb{H}}$.
We add that our basic calculational  result consists in detailed   presentation of $\kappa$-deformed quantum phase space as physically interesting example of  Hopf  bialgebroid (see also \cite{lukx}).

In the paper we introduced the notion of  coproduct  gauges describing  a freedom of coproducts in the Hopf algebroid framework.  For quantum phase spaces the coproduct gauges describe model -- dependent ways of composing the global two--particle phase space from the phase   space coordinates of two its constituents.
 
In a simple case of D=3 nonrelativistic phase space the momenta coproducts describe global 
2--particle momentum as follows
\begin{equation}\label{jlu5.1}
\Delta(p_i) = p_i \otimes 1 + 1 \otimes p_i \quad \longleftrightarrow \quad
p^{(1+2)}_{i} = p^{(1)}_i  + p^{(2)}_i
\end{equation}
and the formula for 2--particle center--of--mass coordinate is given by well--known formula
\begin{equation}\label{jlu5.2}
x^{(1+2)}_i = \frac{m_1}{m_1 + m_2 } \, x^{(1)}_{i} 
+
\frac{m_2}{m_1 + m_2 } \, x^{(2)}_{i} 
\end{equation}
which can be represented  as well by the following one--parameter class of coproduct gauges
$\Lambda_i$ (see also \cite{lukx}, formula (34) in the limit $\kappa \to \infty$)

\begin{equation}\label{jlu5.3}
\begin{array}{l}
\Delta(x_i) = x_i \otimes 1  + \widehat{\Lambda}_i
\\[10pt]
\widehat{\Lambda}_i = (\alpha - 1) (x_i \otimes 1 - 1 \otimes x_i) \qquad \quad
\alpha = \frac{m_1}{m_1 + m_2}\, .
\end{array}
\end{equation}
The coproduct gauges describe different ways of composing center-of-mass coordinate 
$x_i^{(1+2)}$ for different composite systems, characterized by the masses ($m_1, m_2$) of its constituents.  
From this point of view therefore the coproduct gauge freedom 
is not unphysical as in the standard gauge theories, 
but characterizes different choices of dynamical systems leading to  the same phase space algebra (Poisson algebra in classical mechanics, (deformed) Heisenberg algebra in quantum theory).

One can also consider the formulae  describing  $D=4$ relativistic  center of mass coordinates 
$\widehat{x}_\mu^{(1+2)}$ (see e.g. \cite{40,41})  which can be described as the bialgebroid coproduct, with coproduct gauges depending as well on negative powers of energy and three--momenta components.
 In such a way by considering the $(10+10)$ dimensional phase space ${\mathcal{H}}^{(10,10)}$,  containing Lorentz algebra and   the dual Lorentz group parameters,  one can describe in principle various relativistic center-of-mass  coordinates for two particles  with arbitrary spin.

Finally we present an outlook.  We shall point out some ways of extending the presented result, employing the notion of bialgebroid structure:
\vskip .1in

i) The bialgebroid formulae  for various phase space algebra bases.

In this paper (see also \cite{lukx}) we consider the $\kappa$-deformed phase space generated by Majid--Ruegg basis \cite{majidruegg} of corresponding $\kappa$-deformed Poincar\'{e}--Hopf algebra.
 In such a basis the standard $\kappa$-deformed phase space is   described by centrally extended  8--dimensional Lie algebra.
 Such Lie algebraic structure is preserved only under the linear change ($\widehat{p}_\mu \to \widehat{p}{\, }'_\mu = \alpha _\mu^{\ \nu} \widehat{p}_\nu$) of momenta basis.
 The bicrossproduct structure of Poincar\'{e}--Hopf algebra which implies quantum covariance properties of phase space remains  however valid if we change the fourmomenta basis in arbitrary nonlinear way
\begin{equation}\label{jlu5.4}
\widehat{p}_\mu  \to \widehat{p}{\, }'_\mu = F_\mu (\widehat{p}) 
\qquad \qquad  F_\mu (0) = 0 \, .
\end{equation}
In particular one can choose $F_\mu (\widehat{p})$ in  a way leading to classical Poincar\'{e} algebra in algebraic sector of  $\kappa$-deformed Poincar\'{e} algebra 
\cite{33}--\cite{35}.
 In such a way we get more complicated nonabelian formulae 
for the composition law of fourmomenta, 
which unfortunately are not endowed with physical interpretation.
\\[10pt]
ii) Hopf algebroid structure of generalized phase space ${\mathcal{H}}^{(10,10)}$.

In order to describe the global phase coordinates 
for two particles with relativistic spin one should include 
into the phase space algebra 
the Lorentz group parameters
$\widehat{\lambda}^\mu_{\, \nu}$ and dual Lorentz algebra generators 
$\widehat{m}_{\mu\nu}$ (for undeformed case these ideas were considered earlier \cite{17}--\cite{22}). 
 The noncanonical pair of dual generators 
($\widehat{\lambda}_\mu^{\ \nu}, \widehat{m}^\rho_{\ \lambda}$) 
satisfying the relation (17) can be also replaced by a canonical pair 
($\widehat{\omega}^\mu_{\ \nu}, \widehat{\pi}^\rho_{\  \lambda} $) where 
$\widehat{\lambda}^\mu_{\ \nu} = \delta^\mu_{\  \nu} + \widehat{\omega}^\mu_{\ \nu}$  
and ($\eta_{\mu\nu} = diag(-1, 1, 1, 1)$)

\begin{equation}\label{jlu5.5}
\left[ \widehat{\lambda}_\mu^{\ \nu}, \widehat{\omega}^\rho_{\ \lambda} \right] =
 i \left( \delta_\mu^{\ \rho} \, \delta^\nu_{\ \lambda} - \delta _\mu^{\ \lambda} \, \delta^\nu_{\ \rho}
\right)
\end{equation}
consistent with $\widehat{\omega}^\rho_{\ \lambda} = - \widehat{\omega}^\lambda_{\ \rho}$
 and the formula $\widehat{m}^\rho_{\  \lambda} = \widehat{\omega}^\rho_{\ \nu} \, \widehat{\pi}^\nu_{\ \lambda} - \widehat{\omega}^\lambda_{\ \nu} \, \widehat{\pi}^\nu_{\ \rho}$. 
 The calculations of the target map and coproduct gauge freedom for ${\mathcal{H}}^{(10,10)}$ are now  under consideration.
\\[10pt]
iii) Hopf superalgebroid structure of deformed supersymmetric quantum -- deformed phase space.
 
Following the analogy  respecting the  ``sign rules'' between Hopf algebras and Hopf superalgebras one can define the $Z_2$-graded algebraic structure of Hopf superbialgebroids (see e.g.~\cite{karx}).
 In order to extend supersymmetrically the $\kappa$-deformed phase space presented  in this paper one can use already known results about the supersymmetric extension of $\kappa$-deformed Poincar\'{e} and its dual quantum Poincar\'{e} supergroup \cite{13c,Wessx}.
 For such a case one should firstly  employ  the supersymmetric extension of Heisenberg double, in particular for $\kappa$-deformed Poincar\'{e}--Hopf algebra.
\\[10pt]
iv) Twisted bialgebroids and the deformation of  quantum phase spaces.

One can extend the twist deformation technique of Hopf algebras to Hopf algebroids \cite{Xu}.
 In such a way one can obtain from Hopf algebroids with commutative base algebra (e.g. standard Heisenberg algebra) the Hopf algebroid with noncommutative base, which can be effectively calculated in terms of twist factor by so--called $\star$-product formalism.
 In particular one can apply this technique to  dynamical quantum deformations, described by parameter -- dependent classical $r$-matrices satisfying dynamical YB equation 
(for the link of dynamical quantum deformations with Hopf algebroids see  e.g.  \cite{Xu,doninmudrov}).

\begin{acknowledgments}

J.~L. and M.~W. have been supported by 
Polish National Science Center project 2014/13/B/ST2/04043.
\end{acknowledgments}

\end{document}